\documentclass[review]{elsarticle}
\usepackage{multirow}
\usepackage{graphicx}
\usepackage{caption}
\usepackage{lineno,hyperref}
\usepackage{amsmath}
\modulolinenumbers[5]

\usepackage{pifont}
\usepackage{listings}
\newcommand{\cmark}{\ding{51}}%
\newcommand{\xmark}{\ding{55}}%

\journal{Medical Image Analysis}

\begin{document}

\begin{frontmatter}

\title{A Gradient Mapping Guided Explainable Deep Neural Network for Extracapsular Extension Identification in 3D Head and Neck Cancer Computed Tomography Images}
% \tnotetext[mytitlenote]{Fully documented templates are available in the elsarticle package on \href{http://www.ctan.org/tex-archive/macros/latex/contrib/elsarticle}{CTAN}.}

% \address{Radarweg 29, Amsterdam}
% \fntext[myfootnote]{Since 1880.}

% \author[mymainaddress,mysecondaryaddress]{Elsevier Inc}

% \author[mysecondaryaddress]{Global Customer Service\corref{mycorrespondingauthor}}
% \cortext[mycorrespondingauthor]{Corresponding author}

% \address[mymainaddress]{1600 John F Kennedy Boulevard, Philadelphia}
% \address[mysecondaryaddress]{360 Park Avenue South, New York}

\author[1]{Yibin Wang}
\author[1]{Abdur Rahman}
\author[2]{W. Neil. Duggar}
\author[2]{P. Russell Roberts}
\author[2]{Toms V. Thomas}
\author[1]{Linkan Bian}
\author[1]{Haifeng Wang\corref{cor1}}
\cortext[cor1]{Corresponding author. E-mail address: wang@ise.msstate.edu}

\address[1]{Department of Industrial and Systems Engineering, Mississippi State University, Mississippi State, MS, 39762, USA}
\address[2]{Department of Radiation Oncology, University of Mississippi Medical Center, Jackson, MS, 39216, USA}

\begin{abstract}
Diagnosis and treatment management for head and neck squamous cell carcinoma (HNSCC) is guided by routine diagnostic head and neck computed tomography (CT) scans to identify tumor and lymph node features. Extracapsular extension (ECE) is a strong predictor of patients' survival outcomes with HNSCC. It is essential to detect the occurrence of ECE as it changes staging and management for the patients. Current clinical ECE detection relies on visual identification and pathologic confirmation conducted by radiologists. Machine learning (ML)-based ECE diagnosis has shown high potential in the recent years. However, manual annotation of lymph node region is a required data preprocessing step in most of the current ML-based ECE diagnosis studies. In addition, this manual annotation process is time-consuming, labor-intensive, and error-prone. Therefore, in this paper, we propose a Gradient Mapping Guided Explainable Network (GMGENet) framework to perform ECE identification automatically without requiring  annotated lymph node region information. The gradient-weighted class activation mapping (Grad-CAM) technique is proposed to guide the deep learning algorithm to focus on the regions that are highly related to ECE. Informative volumes of interest (VOIs) are extracted without labeled lymph node region information. In evaluation, the proposed method is well-trained and tested using cross validation, achieving test accuracy and AUC of 90.2\% and 91.1\%, respectively. The presence or absence of ECE has been analyzed and correlated with gold standard histopathological findings.
\end{abstract}


\end{frontmatter}

\section{Introduction}

Head and neck squamous cell carcinoma (HNSCC) is one of the most common cancers worldwide, diagnosed in more than 550,000 patients and causing over 300,000 deaths annually \citep{jemal2013annual}. For HNSCC diagnosis and treatment plan selection, CT scans are collected and analyzed to identify clinical tumor and lymph node features \citep{Kann_2018}. In spite of modern imaging techniques, there are certain radiographic features that remain difficult to detect by clinicians, especially the presence of lymph node extracapsular extension (ECE). ECE occurs when metastatic tumor cells within the lymph node break through the nodal capsule into surrounding tissues. It is crucial to identify whether ECE occurs or not for HNSCC patient treatment management. However, current detection in practice mainly relies on the visual identification, including lymph node annotation and pathologic confirmation, which can be extremely labor-intense and time-consuming. Human errors are also inevitable. Therefore, we perform ECE identification automatically using advanced 3D deep learning technique with explainable gradient-based approach, and at the same time, not requiring manual lymph node annotation.

Deep neural networks (DNNs) have demonstrated a great success in many image recognition tasks. Nevertheless, applying neural networks models on high resolution CT scans is very computational extensive. The dimension of CT scans for head and neck cancer (HNC) patients is usually 512$\times$512 with more than a hundred slices. To handle such high resolution input, existing models use a combination of down-sampling, dividing, and/or coarse-to-fine schemes \citep{hou2016patch, chlebus2018deep, vorontsov2018liver}. In this research, we perform a two-step learning scheme. The first step is volumes of interest (VOIs) self-extraction with DNNs to provide explainable insights. The second step will train a classifier with the extracted explainable VOIs. In this way, accurate ECE related lymph node regions are not demanded, and segmentation effort could be saved.

Even though deep learning models have achieved impressive prediction accuracies, their inner non-linear structure makes them highly non-transparent to be explained by human that what information in the input data makes them actually arrive at their decisions \citep{samek2017explainable}. Gradient-weighted class activation mapping (Grad-CAM) is one of the methods that look into the deep learning black boxes. The gradients of the target concept is used, flowing into the final convolutional layer to produce a coarse localization map highlighting the important regions in the image for predicting the concept \citep{selvaraju2017grad}.

In this paper, we propose a novel deep learning framework, Gradient Mapping Guided Explainable Network (GMGENet), to detect ECE from 3D head and neck CT scans without requiring annotated lymph node region information. The task is to classify patients’ CT scans with ECE positive/negative categories and point out the causes in the 3D CT scans. 

Different from the typically ECE detection algorithms, where the annotation of lymph node regions are required, the proposed GMGENet model includes a self-extractor, which can be trained separately to extract the explainable VOIs that are related to the ECE positive class. The self-extractor is designed based on Grad-CAM. Then, a DenseNet classifier is trained based on the VOIs suggested by the self-extractor and tested on an independent test set. Ground truth label has been provided for validating the self-extraction performance, and model explainability analysis are designed in the experiments.
 
The proposed network is trained by the inputs that are highly related to ECE information. Therefore, extracted VOIs will contain explainable features for further classification task. In summary, the contributions of this research are highlighted as follows:

\begin{enumerate}[(1)]
\item We propose a novel deep learning framework GMGENet with two-step learning scheme for ECE identification in head and neck cancer. The proposed architecture includes a self-extractor, which can extract the explainable VOIs that are related to the class ECE positive. The proposed GMGENet has achieved better performance than conventional networks.
\item We train and test our proposed model on a real-world collected dataset. Lymph node label is not required in the classifier, which will promote the implementation as well as improve the efficiency of artificial intelligence-assist ECE detection.
\item To further illustrate ECE identification performance, we apply a gradient-based mapping approach to generate 3D ECE probability heatmaps to provide visualization during the training. This technique will indicate the important regions related to ECE and increase the model explainability.
\end{enumerate}

The structure of this paper is organized as following. Section 2 provides an overview of related algorithms and applications in head and neck ECE identification. Section 3 describes the details of our proposed explainable GMGENet model for ECE classification. Data preparation and experimental results are illustrated in Section 4. The research findings and future work are concluded in Section 5.

\section{Related Works}
\subsection{Head and Neck Cancer Diagnosis}
In recent years, studies have been conducted with deep learning models on ECE identification in head and neck cancer. Kann er al. \citep{kann2018pretreatment} has trained a 3D convolutional neural network (CNN) with annotated lymph node samples and correlated pathology labels and achieved an area under the curve (AUC) of 91\%. In another study Kann et al. \citep{kann2020multi} proposed a deep learning algorithm to identify the extranodal extension (ENE) in HNSCC. They used CT scans with lymph node annotations for ENE-positive and ENE-negative classes. In their work, they compared the performance of their proposed method with the performance of two neuroradiologists and found that their proposed method outperforms the neuroradiologists. Ho et al. \citep{ho2020classifying}  studied a multilayer perceptron with feature extraction to identify ENE and non-ENE lymph nodes and they achieved 77\% accuracy. Ariji et al. \citep{ariji2019contrast} used contrast-enhanced segmented CT images and a deep learning model to classify the lymph node metastasis. They achieved a result of 78.2\% accuracy and 80\% AUC. In another study, Ariji et al. \citep{ariji2021automatic} applied an object detection model to automatically detect the lymph nodes in patients with oral squamous cell carcinoma (a type of HNSCC) and achieved a performance level of 73\% and 52.5\% recall for metastasis and non-metastasis lymph nodes respectively. Tomita et al. \citep{tomita2021deep} proposed Xception-based lymph node classification method with contrast-enhanced CT images. Chen et al. \citep{chen2019combining} proposed a hybrid method combining many-objective radiomics and 3D CNN to predict lymph node metastasis. They concluded that the hybrid method outperformed both of the stand-alone radiomics model and CNN model. All these works mainly focus on ECE classification on lymph node annotated CT image data. This lymph node annotation requires pathologic confirmation and visual identification by expert radiologists. Therefore, there is a need of identifying the VOI automatically . To generate clinical target volumes (CTV) for lymph nodes, Cardenas et al. \citep{cardenas2021generating} used a U-Net auto segmentation model. However, this doesn't address the issue of generating the VOIs automatically because the U-Net was still trained with lymph node annotated CT images. Moreover, they haven't used these images for ECE classification. Therefore, existing studies show that lymph node annotated CTs are required for training ECE diagnosis machine learning models, which is a key limitation of current computer-aided ECE diagnosis models. To overcome the limitation, which typically involves time consuming manual processes,  a GMGENet model is proposed in this study. A summary of related works and the comparsion with our proposed GMGENet model  can be found in Table \ref{related works}.

\begin{table}[!htp]
\centering
\caption{Summary of related works on ECE identification in HNSCC. $\dagger$ Accuracy, $\displaystyle \ast$ Recall, $+$ Positive ECE, $-$ Negative ECE, I-II,I,II Three Stages of Lymph Node Metastasis. PET = Positron Emission Tomography, CT = Computed Tomography, and MRI = Magnetic Resonance Imaging}
\label{related works}
\resizebox{1\textwidth}{!}{%
\begin{tabular}{@{}cccccccc@{}}
\hline
Year &
  Model &
  \begin{tabular}[c]{@{}c@{}}Dataset \\ Image\\ Type\end{tabular} &
  \begin{tabular}[c]{@{}c@{}}Number of \\ Patients in \\ the Dataset\end{tabular} &
  \begin{tabular}[c]{@{}c@{}}LN Mask\\ Required\end{tabular} &
  \begin{tabular}[c]{@{}c@{}}Model \\ Explain-\\ ability\end{tabular} &
  AUC &
  References \\ \hline
2018 &
  DualNet &
  CT &
  270 &
  \cmark &
 \xmark &
  0.65 – 0.69 &
  \citep{kann2018pretreatment} \\
2019 &
  DualNet &
  CT &
  82 &
  \cmark &
  \xmark &
  0.84 &
  \citep{kann2020multi} \\
2019 &
  AlexNet &
  CT &
  45 &
  \cmark &
  \xmark &
  0.80 &
  \citep{ariji2019contrast} \\
2019 &
  Hybrid &
  \begin{tabular}[c]{@{}c@{}}PET \& \\ CT\end{tabular} &
  59 &
  \cmark &
  \xmark &
  0.88 &
  \citep{chen2019combining} \\
2020 &
  MLP &
  MRI &
  25 &
  \cmark &
 \xmark &
  $0.77^{\dagger}$ &
  \citep{ho2020classifying} \\
2021 &
  DetectNet &
  CT &
  56 &
  \cmark &
 \xmark &
  \begin{tabular}[c]{@{}c@{}}$0.73_+^{\displaystyle \ast }$,\\ $0.52_-^{\displaystyle \ast }$\end{tabular} &
  \citep{ariji2021automatic} \\
2021 &
  Xception &
  CT &
  39 &
  \cmark &
  \xmark &
  \begin{tabular}[c]{@{}c@{}}$0.85_{I-II}$,\\ $0.80_{I}$,\\ $0.90_{II}$\end{tabular} &
  \citep{tomita2021deep} \\
\multicolumn{2}{c}{Proposed GMGENet}&
  CT &
  130 &
  \xmark &
  \cmark &
  0.91 &
  The proposed work \\ \hline
\end{tabular}%
}
\end{table}
%\end{landscape}
% \usepackage{booktabs}
\begin{table}[h]
\centering
\caption{Summary of model explainability methods}
\label{related works2}
\resizebox{1\textwidth}{!}{%
\begin{tabular}{@{}ll@{}}
\hline
Method               & Key Characteristics                                                                                                                                      \\ \hline
Saliency Map \citep{simonyan2013deep}         & \begin{tabular}[c]{@{}l@{}}Can't distinguish between positive and negative information \\ and creates visually noisy map\end{tabular}           \\
DeConvNet \citep{zeiler2014visualizing}           & \begin{tabular}[c]{@{}l@{}}Can be only used with CNN models based on ReLU \\ activation\end{tabular}                                            \\
GBP \citep{springenberg2014striving}                 & \begin{tabular}[c]{@{}l@{}}Combination of saliency map and DeConvNet, limited to \\ CNN models with ReLU activations\end{tabular}               \\
SmoothGrad \citep{smilkov2017smoothgrad} &
  \begin{tabular}[c]{@{}l@{}}Reduces the noise but inherits the sensitivity of the \\ underlying attribution method and is dependent on chosen \\ hyperparameters\end{tabular} \\
Gradient$\odot$Input \citep{shrikumar2016not} & \begin{tabular}[c]{@{}l@{}}Improves the sharpness of attribution maps but is sensitive \\ to input transformation\end{tabular}                  \\
IG \citep{sundararajan2017axiomatic}                  & Dependent on input baseline                                                                                                                     \\
DeepLIFT \citep{shrikumar2017learning}             & \begin{tabular}[c]{@{}l@{}}Can't be applied to models involving multiplicative rules,  \\ and is dependent on input baseline\end{tabular}       \\
GradCAM \citep{selvaraju2017grad}             & \begin{tabular}[c]{@{}l@{}}Produces class-specific gradient information and localization \\ map of important features of the input\end{tabular} \\ \hline
\end{tabular}%
}
\end{table}

\subsection{Model Explainability Methods}
In recent years, the studies conducted on explainability aim to identify the features that mostly influence the decision of a model \citep{singh2020explainable}. Two broad types of explainability methods have been recognized \citep{nielsen2021robust}: (1) feature perturbation based, and (2) gradient based. The feature perturbation based explainability methods refers to perturbing the input features by slightly changing their values, and then observing the effects of these perturbation on the model performance. The gradient based methods works in two steps. First, the gradients of the output with respect to input or the extracted features are calculated through back-propagation. Second, these gradients are used to estimate the attribution score. In the gradient based methods, these attribution scores indicate how much each feature is contributing to the predictions of the model. The gradient based methods are more robust to input perturbations compared to the perturbation based methods \citep{nielsen2021robust}. Therefore, when robustness is the priority, the gradient-based methods are more preferred. Most widely used gradient based explainability methods are Saliency Map \citep{simonyan2013deep}, DeConvNet \citep{zeiler2014visualizing}, Guided Backpropagation (GBP) \citep{springenberg2014striving}, SmoothGrad \citep{smilkov2017smoothgrad}, Gradient $\odot$ Input \citep{shrikumar2016not} ($\odot$ represents multiplication), Integrated Gradients (IG) \citep{sundararajan2017axiomatic}, Deep Learning Important FeaTures (DeepLIFT) \citep{shrikumar2017learning}, and Gradient-weighted Class Activation Mapping (GradCAM) \citep{selvaraju2017grad}. While choosing the suitable method for explainability, one should consider the limitations of these methods. For instance, the major drawback of Saliency Map is that they are visually noisy. DeConvNet is limited to ReLU activation only. While GBP combines the operations of both Saliency Map and DeConvNet, it shows the drawbacks of both. SmoothGrad reduces the noise by calculating and averaging the Saliency Map, but it inherits the sensitivity of the underlying attribution method \citep{inbook}. DeepLIFT is not applicable to models involving multiplicative rules \citep{singh2020explainable}. An ideal attribution map should be insensitive to input transformations and model hyperparameters, and should not be dependent on input baseline. Gradient $\odot$ Input is sensitive to input transformation and DeepLIFT and IG are dependent on input baseline. Moreover, SmoothGrad is dependent on chosen hyperparameters. While the above mentioned methods are limited by their drawbacks, GradCAM can overcome these limitations and generate more explainable attribution maps. GradCAM uses class-specific gradient information flowing into the final convolution layer of a CNN and produce a localization map of important features of the input \citep{selvaraju2017grad}. To put it another way, GradCAM analyzes which regions are activated in the feature maps of the last convolutional layer.  \citep{nielsen2021robust}. Therefore, we chose GradCAM as the model explainability method for this study. A summary of model explainability methods  can be found in Table \ref{related works2}.

\section{Methodology}

\subsection{Proposed GMGENet Model}

There has been extensive research regarding 2D CNN architectures to optimize performance for 2D image analysis, and several successful architectures, for instance, AlexNet, VGGNet, ResNet, and DenseNet have emerged over the past few years \citep{krizhevsky2012imagenet, simonyan2014very, he2016deep, huang2017densely}. Nevertheless, the adaptation of these models to 3D image analysis problems, especially CT-based ECE classifcation, has been less established. 

\begin{figure}[h]
    \centering
    \includegraphics[width=\linewidth]{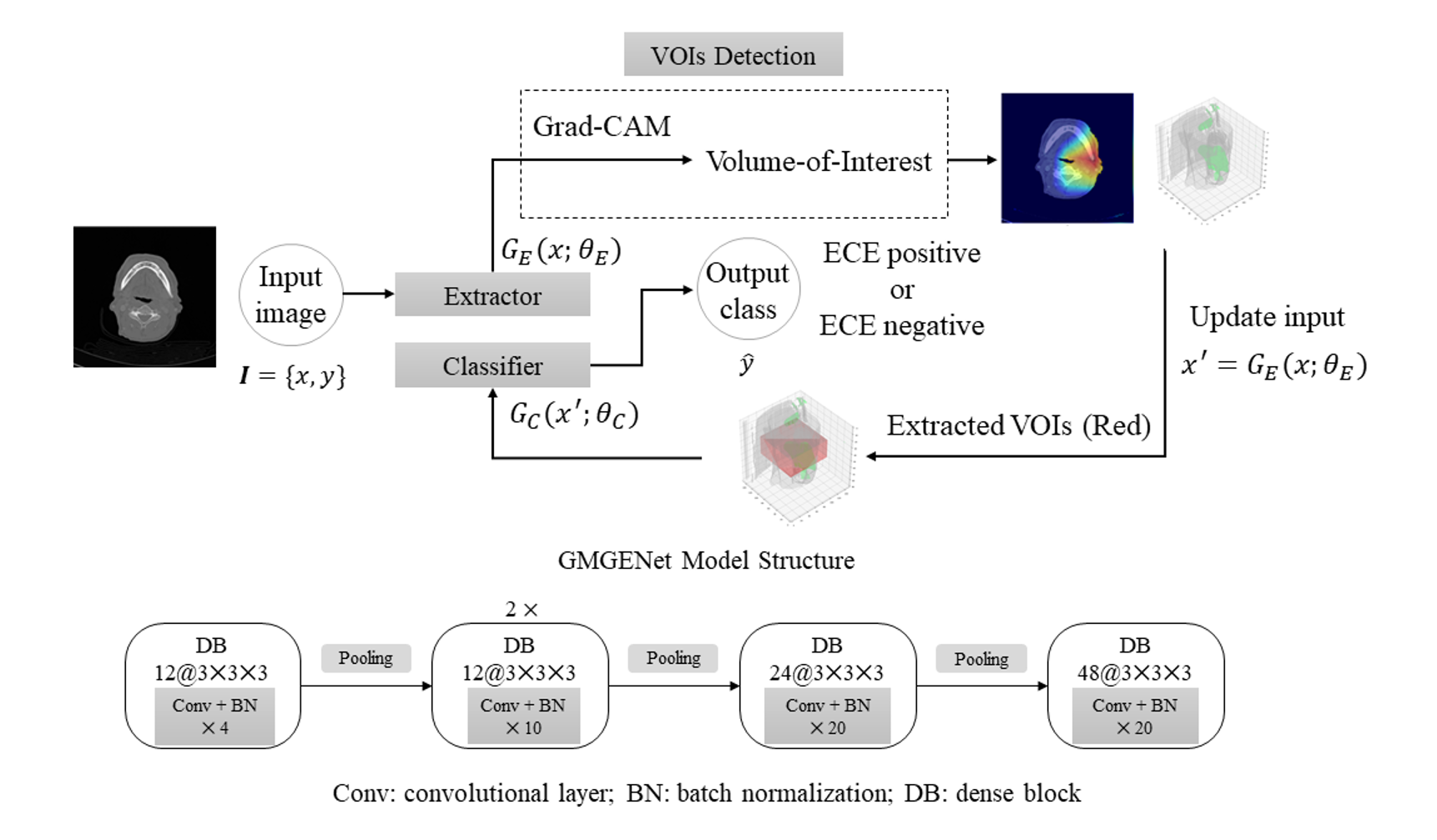}
    \caption{The proposed GMGENet model framework.}
    \label{Fig1}
\end{figure}

The architecture of the proposed GMGENet framework is shown in Figure \ref{Fig1}. We denote input CT image dataset is referred as $\textit{\textbf{I}} = \left \{x,y \right \}$, where $x$ is the original input CT scan images, and $y \in \left \{0,1 \right \}$ is the class label, where 0 and 1 denote ECE negative and positive, respectively. There are two main components in the GMGENet framework: 1) a VOI self-extractor $G_{E}$, and 2) an ECE classifier $G_{C}$, with the parameters of $\theta_{E}$, $\theta_{C}$ for each mapping, respectively. The mapping functions are gathered as Equations (\ref{eq1})-(\ref{eq2}):

\begin{equation}
    x' = G_{E}(x; \theta_{E}) \label{eq1}
\end{equation}
\begin{equation}
    \hat{y} = G_{C}(x'; \theta_{C}) \label{eq2}
\end{equation}

where $x'$ is the updated VOI inputs ($x' \subset x$), and $\hat{y}$ is the predicted class vector. The aim of the self-extractor is to extract VOIs from the original input CT scans. It is noted that the original inputs also follow some basic preprocessing steps. The details are demonstrated in the data preparation and preprocessing subsection. 

During the training, important regions that the self-extractor considers related to ECE can be highlighted and extracted. This extractor is built based on 3D DenseNet architecture and VOI extraction is accomplished through Grad-CAM. The 2D and 3D view of a extracted sample has been shown in Figure \ref{Fig1}. The extracted VOIs will be used to train the classifier for ECE classification. Additionally, both the self-extractor and classifier are based our previous studies of a 3D DenseNet Model \citep{wang2021extracapsular}, in which the convolutional layers are with progressively increasing number of features and fully connected layers. 3D DenseNet architecture is also shown in Figure \ref{Fig1}. There are five dense blocks where all layers with matching feature-map sizes are connected directly with each other in a module. The loss function of this network can be represented as:

\begin{equation}
    E(\theta_{E}, \theta_{C})=\sum_{i=1,2,...,N}L_{y}(G_{C}(G_{E}(x_{i};\theta_{E});\theta_{C} )y_{i}) = \sum_{i=1,2,...,N}L_{y}^{i}(\theta_{E}, \theta_{C})
\end{equation}

where $L_{y}$ refers to the loss of the classifier, and $L_{y}^{i}$ represents the corresponding loss function evaluated at the \textit{i}-th training sample. The parameters $\hat{\theta}_{E}, \hat{\theta}_{C}$ are optimized using Adadelta optimizer \citep{zeiler2012adadelta} by minimizing the classifier loss based on the equation:

\begin{equation}
    (\hat{\theta}_{E}, \hat{\theta}_{C}) = \operatorname*{argmin}_{\theta_{E}, \theta_{C}} E(\theta_{E}, \theta_{C})
\end{equation}

The CNN core in our proposed GMGENet is 3D DenseNet. DenseNet is first proposed in \citep{huang2017densely} that connects each layer to every other layer in a feed-forward fashion. Consider a single image input $x_{0}$ passes through a deep neural network with $L$ layers. The layer $l$ applies a nonlinear operation $F_{l}\left ( \cdot \right )$ such as convolution and batch normalization to the given input. Hence, conventional feed-forward networks will give such transition: $x_{l} = F_{l}\left ( x_{l-1} \right )$. Following the densely connectivity pattern, direct connections are built to all subsequent layers in each dense block:

\begin{equation}
    x_{l} = F_{l}\left ( \left [ x_{l-n}, x_{l-n+1}, ..., x_{l-1} \right ] \right )
\end{equation}

where n is the number of layers in a dense block, and $x_{l-n}, ..., x_{l-1}$ refers to the concatenation of the feature maps produced in all the layers in a dense block. We implemented this basic densely-connected architecture to handle 3D images for both extractor and classifier networks.

\subsection{Model Explainability}

Conventional class activation mapping (CAM) \citep{zhou2016learning} approach reforms CNN architectures in image classification missions by replacing fully-connected layers with convolutional and global average pooling layers to achieve class-specific feature maps. Grad-CAM further uses the gradient information flowing into the last convolutional layer of CNNs to understand the importance of each neuron for a decision of interest. Because extracted convolutional features naturally retain spatial information which is mostly lost in fully-connected layers, it is generally considered that the last convolutional layers are expected to have the best trade-off between high-level semantics and detailed spatial information. The neurons in these layers are optimized for particular class-specific information in the image, and Grad-CAM localization heatmap can be generated to represent where the model concentrates more on to make the particular decision.

In terms of the class localization heatmap generation, suppose for one category $c$, the target heatmap has the dimension of $R^{m \times n}$, where $m$ and $n$ refer to width and height, respectively. The score of the class $c$ is collected as $y_{c}$. Thus, the gradient of the feature maps activation $A^{k}$ can be expressed as, 

\begin{equation}
    g_{c}(A^{k})=\frac{\partial y_{c}}{\partial A^{k}}
\end{equation}

where $k$ refers to the channel index. The gradients are further averaged as the neural importance weight $\alpha_{k}^{c}$ in each channel over the width and height dimensions by,

\begin{equation}
    \alpha_{c}^{k}=\frac{1}{Z} \sum_{i} \sum_{j} \frac{\partial y_{c}}{\partial A^{k}_{i,j}}
\end{equation}

where Z is the spatial resolution of the k feature map with spatial index $(i,j)$. This updated neural importance weight $\alpha_{k}^{c}$ is named grad-weight that represents a partial linearization of the DNN downstream from $A$, and captures the importance of feature map $k$ for a target class $c$. Then, Grad-CAM is a weighted sum of feature maps, followed by a ReLU operation as,

\begin{equation}
    H_{Grad-CAM}^{c}=RELU(\sum_{k}\alpha^{c}_{k}A^{k})
\end{equation}

The ReLU is applied to the linear combination $\sum_{k}\alpha^{c}_{k}A^{k}$ of maps because only the features that have a positive influence on the class are interesting to explore, where the pixel intensity should be increased for a larger $y_{c}$. Negative pixels are likely to relate to other classes in the classification, and more accurate visual localization heatmaps are obtained.

\section{Experimental Results}

\subsection{Data Preparation and Preprocessing}

The HNSCC dataset used in this study are collected by the University of Mississippi Medical Center. HNC patients were retrospectively reviewed with the diagnosis of HNSCC that were treated at the institution between 2008 and 2014. Initial staging at diagnosis was performed utilizing CT neck with contrast. We trained and tested our proposed deep learning model based on 130 selected patient data, 80 ECE positive patients and 50 ECE negative patients with the CT scans. A summary of some key features of the dataset have been shown in Table \ref{dataset}.

\begin{table}[!htp]
\centering
\caption{Dataset description. (M: male; F: female; LN: lymph node; SD: standard deviation)}
\label{dataset}
\resizebox{0.75\textwidth}{!}{%
\begin{tabular}{cccc}
\hline
Features      & ECE positive  & ECE negative & Total        \\ \hline
Patient amount        & 80            & 50           & 130          \\
Age (SD)      & 59.05 (10.11) & 57.39 (8.37) & 58.22 (9.73) \\
Gender        & 62 M; 18 F    & 36 M; 14 F   & 98 M; 32 F   \\
LN size (SD)  & 2.66 (1.62)   & 1.75 (0.95)  & 2.21 (1.47)  \\ \hline
\end{tabular}%
}
\end{table}

\begin{figure}[h]
  \centering
  \includegraphics[width=0.8\linewidth]{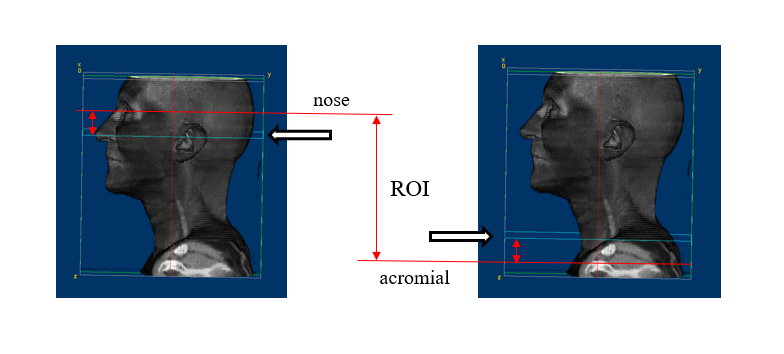}
  \caption{Selected VOIs for ECE classification \citep{wang20203d}.}
  \label{VOI}
\end{figure}

In our data prepossessing procedure, we first applied VOI segmentation to narrow the entire CT scans down to a few particular regions where lymph nodes may locate, so as to remove irrelevant background that will influence the classification accuracy. Hounsfield unit (HU) is widely used in CT scanning to represent values in a standardized and convenient form. Generally, the range of data in the brain and facial tissues are within -100 to 300 HU, excluding the skull, other bones, and calcification’s \citep{muschelli2019recommendations}. We selected the range of -400 to 400 HU recommended by radiologists. After applying the threshold, the bones are excluded and facial tissues are retained. Then the slices of nose and acromial can be identified in the CT scans for VOI segmentation. We selected VOIs based on the slice of nose 3 centimeters upward and the slice of acromial 3 centimeters downward. Figure \ref{VOI} demonstrates the segmented VOIs \citep{wang20203d, wang2021extracapsular}.

\begin{figure}[!b]
  \centering
  \includegraphics[width=0.95\linewidth]{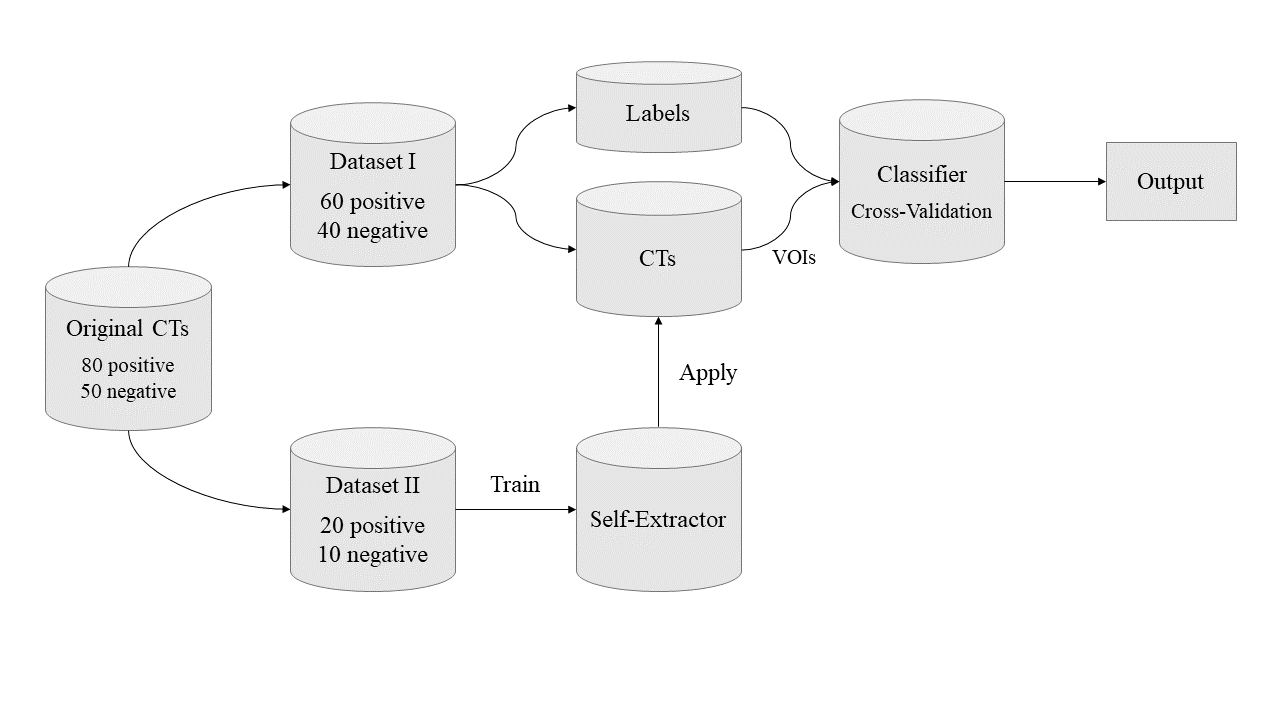}
  \caption{Flow chart of data split scheme}
  \label{Data flow}
\end{figure}

The original CT scan image size is 512 $\times$ 512 for one slice. The number of slices differ from each patient. To further handle the extracted VOIs, we resampled each voxel to represent 1 $\times$ 1 $\times$ 1 millimeter, keeping the same actual spacing. HU values are rescaled between 0 and 1 after clipping beyond -400 and 400 threshold. Then, the input CT scan images were resized to 150 $\times$ 150 $\times$ 90 for the first CNN model. After appling Grad-CAM, a VOI size of 90 $\times$ 90 $\times$ 25 has been extracted for the classifier for classification.

80\%/20\% train-test split has been applied to the dataset, and five-fold cross-validation was used to attain more generalized model and convincing results. It is important to note that in order to avoid data leakage during the self-extractor training, a separate set containing 30 patients has been randomly selected from the original dataset for training the extractor, and this training set was not included in the classifier training and testing. Thus, in terms of model testing, samples are flowed through the self-extractor to get the updated VOIs which can be directly tested by the classifier. The data split flow chart in shown in Figure \ref{Data flow}. ECE related node contours of 5 patients have been provided for Grad-CAM performance measurement with regard to ground truth lymph node label. The output of the first CNN was considered as VOIs that were highly related to ECE. Therefore, the classifier does not require for accurate ECE related lymph node region label.

\subsection{Classification Performance}

We collected the classification performance comparing GMGENet with other two typical deep learning frameworks. A set of averaged results including training and test accuracy as well as other classification measurement metrics have been shown in Table \ref{table1}. Five-fold cross validation has been applied. The standard deviations are also provided. Different classification performance measurements are presented, including accuracy, AUC, sensitivity, and specificity. Accuracy is the proportion of true results, either true positive or true negative, among the test. Sensitivity is the proportion of true positives that are correctly identified, which shows how well the test is at detecting ECE. Specificity is the proportion of the true negatives correctly identified, which suggests how well the test is at identifying normal (ECE negative) condition. These performance metrics can be formulated as:

\begin{equation}
    Accuracy=\frac{TP+TN}{TP+TN+FP+FN}
\end{equation}
\begin{equation}
    Sensitivity=\frac{TP}{TP+FN}
\end{equation}
\begin{equation}
    Specificity=\frac{TN}{TN+FP}
\end{equation}

where true positive (TP) refers to the number of cases correctly identified as ECE positive, false positive (FP) refers to the number of cases incorrectly identified as ECE positive, true negative (TN) refers to the number of cases correctly identified as ECE negative, and false negative (FN) refers to the number of cases incorrectly identified as ECE negative.

\begin{table}[!htp]
\centering
\caption{Test classification performance comparison of different CNN models.}
\label{table1}
\captionsetup{font=scriptsize}
\resizebox{0.85\textwidth}{!}{%
\begin{tabular}{cccc}
\hline
Performance metric & Baseline CNN  & DenseNet      & GMGENet       \\ \hline
Accuracy           & 0.528 (0.029) & 0.712 (0.010) & \textbf{0.902 (0.020)} \\
AUC                & 0.607 (0.002) & 0.740 (0.012) & \textbf{0.911 (0.011)} \\
Sensitivity        & 0.681 (0.001) & 0.696 (0.007) & \textbf{0.909 (0.023)} \\
Specificity        & 0.652 (0.004) & 0.729 (0.006) & \textbf{0.895 (0.009)} \\ \hline
\end{tabular}%
}
\end{table}

\begin{figure}[!htp]
    \centering
    \includegraphics[width=0.85\linewidth]{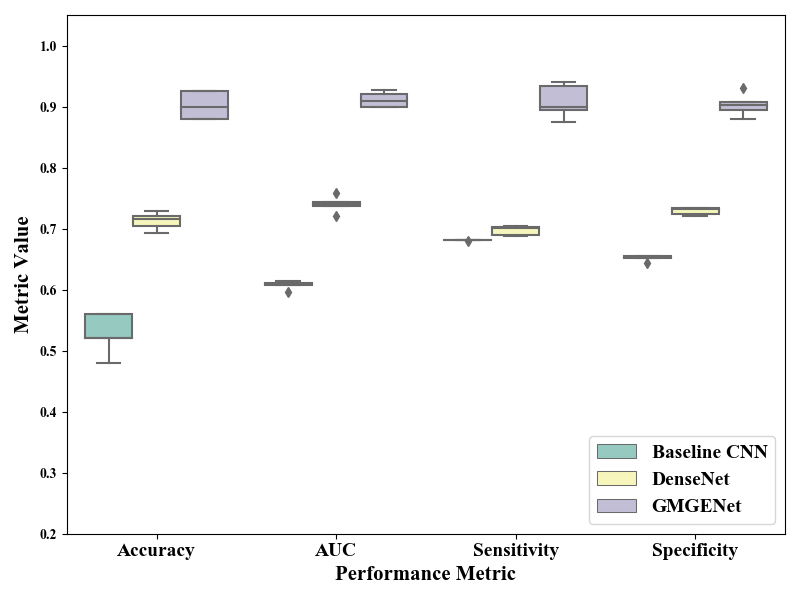}
    \caption{Boxplots of different CNN model test performance}
    \label{Boxplot}
\end{figure}

\begin{figure}[!htp]
  \centering
  \includegraphics[width=0.8\linewidth]{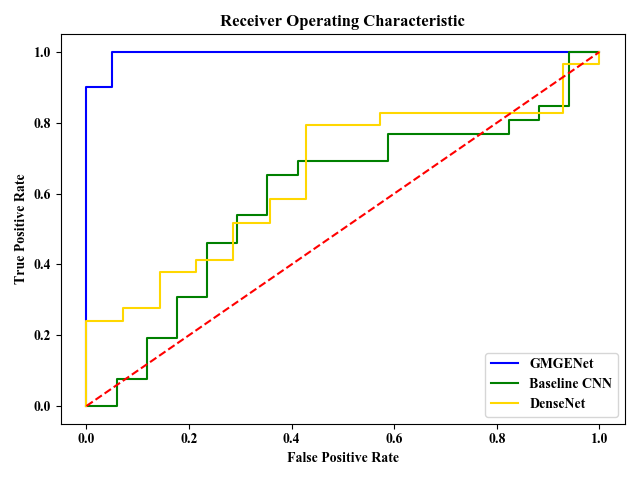}
  \caption{Comparison of different model ROC curves}
  \label{ROC}
\end{figure}

From the test results, we demonstrate that the proposed GMGENet with two-step learning scheme outperforms other conventional 3D CNN Models. 
A simple 3D CNN baseline model with few convolutional layers has been compared \citep{zunair2020uniformizing}. Typical 3D DenseNet was also included in the testing. The 3D baseline model achieved 60.7\% test AUC and 82.5\% accuracy. 3D DenseNet has obtained 74\% test AUC and 71.2\% accuracy. The accuracy of GMGENet improved a lot after the dimension being narrowed down during the self-extractor training, and more accurate VOIs were studied in particular. Training with the updated input, the GMGENet classifier has achieved 91\% AUC and an accuracy of 90.2\%. The ROC curves of different models are shown in Figure \ref{ROC}. The classification results strongly indicate that the proposed GMGENet is capable of extracting valuable information from the original images. Detailed boxplot comparison of each model is presented in Figure \ref{Boxplot}. Even though the standard deviations were a bit higher, all the performance metrics significantly outperformed other models. Guided by Grad-CAM approach, more ECE related features can be learned which is helpful for ECE identification. 

\begin{table}[htp]
\centering
\caption{Paired t-test results of accuracy}
\label{t_test}
\resizebox{0.6\textwidth}{!}{%
\begin{tabular}{cccc}
\hline
Model        & Baseline CNN & DenseNet & GMGENet \\ \hline
Baseline CNN &             & +        & +       \\
DenseNet     &             &          & +       \\
GMGENet      &             &          &         \\ \hline
\end{tabular}%
}
\end{table}

We further performed paired t-test for accuracy. Table \ref{t_test} indicates the statistical test results for test accuracy. With regard to the paired t test, the null hypothesis is that the pairwise difference between the two tests is equal, and the $H_{a}$ is that the difference is not equal to zero. The significance level $\alpha$ is selected as 0.05. The negative $(-)$ signs indicate fail to reject $H_{0}$, and the positive ($+$) signs indicate to reject $H_{0}$. The test results show that the proposed GMGENet model significantly outperforms the baseline model and DenseNet model in terms of accuracy.

\begin{figure}[h]
  \centering
  \includegraphics[width=0.85\linewidth]{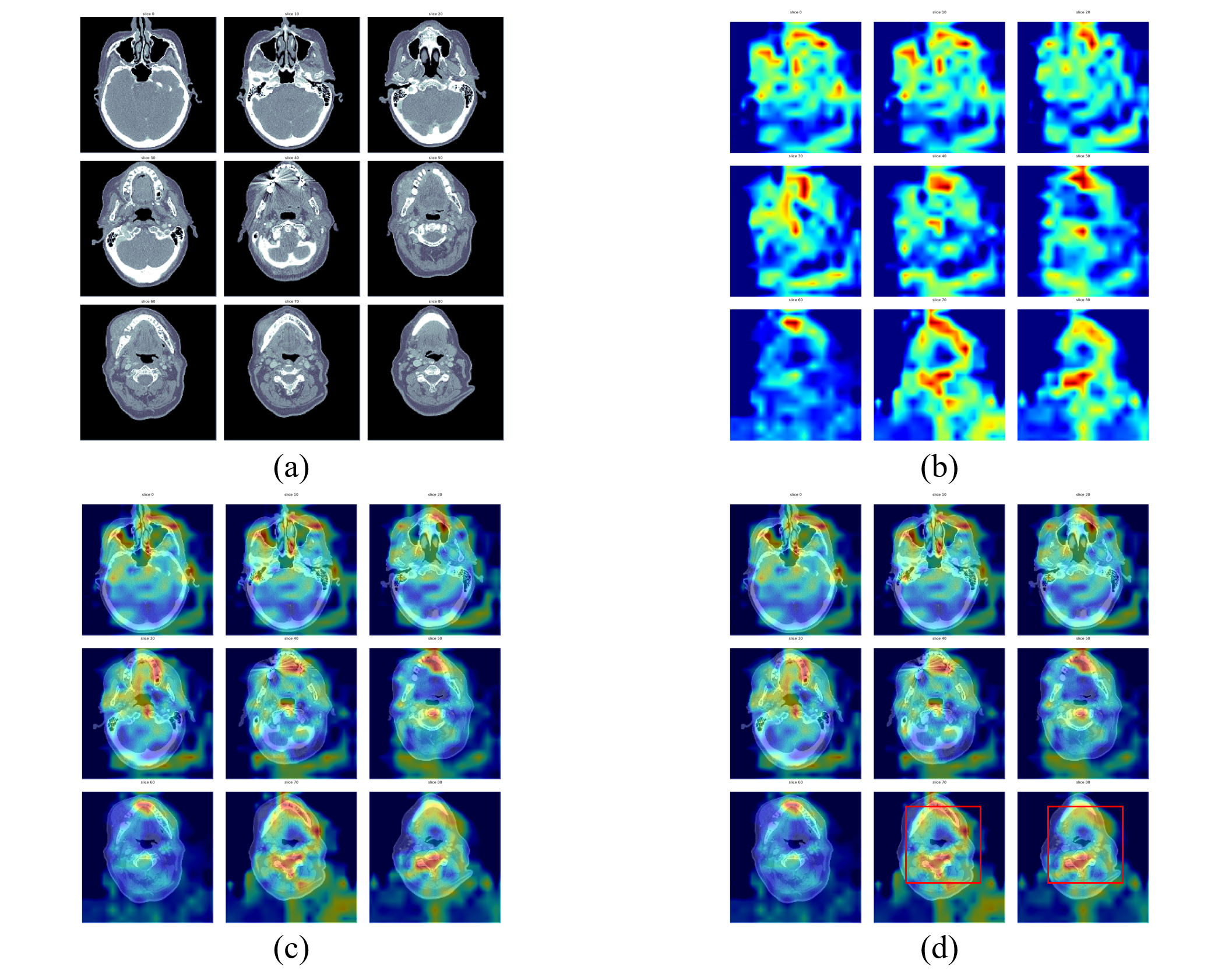}
  \caption{ECE probability heatmaps (plotted every 10 slices). (a) Original CT scan images; (b) Corresponded Grad-CAM heatmaps; (c) Overlapped representations; (d) Overlapped representations with the extracted VOIs}
  \label{Grad-CAM1}
\end{figure}

\subsection{Discussion on Model Explainability}
\begin{figure}[!htp]
  \centering
  \includegraphics[width=0.85\linewidth]{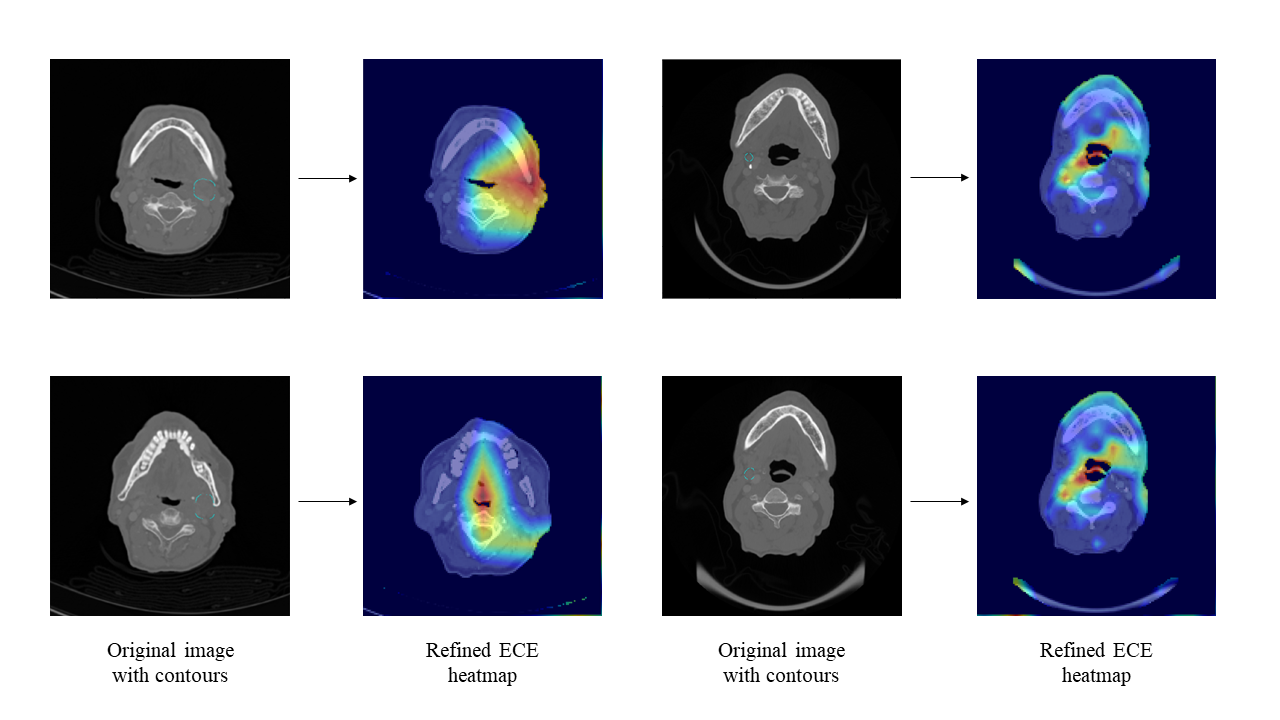}
  \caption{Comparison of original images with lymph node contours and generated ECE heatmaps.}
  \label{Grad-CAM2}
\end{figure}

We evaluated our model explainability through comparing the validated ECE related lymph node contours with the regions that Grad-CAM extracted. Grad-CAM heatmaps on ECE positive class can be visualized in two-dimensional of each slice. Since resampling and resize operations have been conducted in the image preprocessing, the dimensions are narrowed and less slices were available for visualization, compared with the original CTs. In Figure \ref{Grad-CAM1}, we showed one patient sample with its generated ECE probability heatmaps. Every 10 slices have been selected to plot. The extracted are indicated by the red bounding boxes. Because it was noticed that some of regions that are certainly not related to ECE have also been assigned with positive value, we further refined the heatmaps excluding the values out of the head and neck region. Then, we compared the refined heatmaps with the ECE related lymph node regions contoured by radiologists in Figure \ref{Grad-CAM2}. 

In Figure \ref{Grad-CAM2}, two selected patients with original CT scan images and refined heatmaps are shown and illustrated. It can be identified from the drawn contours that the first patient (left column) had ECE occurred one the left side (image right side) lymph node areas, and the second patient had on the right side (image left side) lymph node areas. The heatmaps performed well on identifying the ECE related regions, and the rough ECE corresponded areas have been highlighted. In addition to the lymph node regions, the model also indicated primary tumor area, which may demonstrate further relationship between the extension and the primary tumor. Overall, the Grad-CAM is capable of revealing rough important regions that are related to ECE occurrence, and consequently, those meaningful VOIs can be extracted to train through the second CNN for further classification.

\section{Conclusions}

In this paper, we propose an explainable deep neural Network framework GMGENet, to automatically perform ECE identification without manual lymph node annotation. Grad-CAM approach is proposed to guide the deep learning algorithm to concentrate on the regions that have high probabilities to be related to ECE occurrence. VOIs are extracted through the self-extractor, and the ECE classifier is trained without pre-annotated ECE related lymph node information. In our model evaluation, the proposed GMGENet is trained and evaluated with cross validation. A test accuracy and AUC of 90.2\% and 91.1\% have been achieved, respectively. Furthermore, the self-extractor can be explained through the Grad-CAM results which have shown that ECE related areas are highlighted and extracted. The GMGENet tends to learning meaningful patterns that ECE patients possess. This research also demonstrates the capability to use artificial intelligence for ECE identification. By introducing the possibility heatmap, we can identify where ECE are possible to occur. Therefore, this will contribute to the future clinical implementations of the proposed model, and revealed the unknown features to the radiologists. The outcome of this study is expected to promote the implementation of interpretable artificial intelligence-assist ECE detection.

Further studies will focus on enhancing the model with more validation and testing on larger public datasets. Training on more unseen data will enhance the robustness of the extractor and classifier. From the experimental setting prospective, since currently the ECE positive category has drawn more attention in which Grad-CAM has been implemented for VOIs extraction, more discussion should be added in terms of the ECE negative class. Interesting unknown features in ECE negative patients can be discussed and concluded. In addition, extracted VOIs should be referred to radiologists for more insights. Current VOIs are roughly extracted, accurate and precise detection are worthy to explore and achieve with regard to the clinical ECE identification.

\biboptions{authoryear}
\bibliography{mybibfile}

\end{document}